\documentclass[aps,prb,twocolumn,a4paper,showpacs,superscriptaddress,floatfix,10pt]{revtex4-1}

\usepackage{graphicx}
\usepackage{dcolumn}
\usepackage{bm}
\usepackage{hyperref}
\usepackage[mathlines]{lineno}
\usepackage{amsmath} 
\usepackage{newtxtext,newtxmath} 
\usepackage{comment}
\usepackage{bbm}
\usepackage{epsfig}
\usepackage{helvet}
\usepackage{epstopdf}
\usepackage{dblfloatfix}
\usepackage{multirow}
\usepackage{textcomp}

\begin{document}


\title{Improved Wavelets for Image Compression from Unitary Circuits}

\author{James C. McCord}
\author{Glen Evenbly}
\affiliation{School of Physics, Georgia Institute of Technology, Atlanta, GA 30332, USA}
\date{\today}

\begin{abstract}
We benchmark the efficacy of several novel orthogonal, symmetric, dilation-3 wavelets, derived from a unitary circuit based construction\cite{Evenbly2018}, towards image compression. The performance of these wavelets is compared across several photo databases against the CDF-9/7 wavelets in terms of the minimum number of non-zero wavelet coefficients needed to obtain a specified image quality, as measured by the multi-scale structural similarity index (MS-SSIM). The new wavelets are found to consistently offer better compression efficiency than the CDF-9/7 wavelets across a broad range of image resolutions and quality requirements, averaging $7$-$8\%$ improved compression efficiency on high-resolution photo images when high-quality (MS-SSIM = 0.99) is required.  \end{abstract}
\maketitle

\section{Introduction}
\label{sec:introduction}
The advent and proliferation of compact orthogonal wavelets and discrete wavelet transforms (DWTs) in late 80's and early 90's represented a breakthrough advance in signal analysis\cite{C.1993TenWavelets.,Resnikoff1998WaveletAnalysis,Rao1999WaveletApplications,Walnut2004AnAnalysis,Walker2008AApplications}. Given a set of smoothly varying set of spatially or temporally correlated data, perhaps with a limited content of discontinuities or sharply varying features, DWTs can typically be employed to obtain a highly compressed representation of the data. Consequently wavelets have found significant applications in image, audio and video compression\cite{Smith1990Analysis/synthesisCoding,Bradley1994Wavelet/scalarImages,Villasenor1995WaveletCompression,Bhaskaran1997ImageArchitectures,Vetterli2001WaveletsCompression,Majumder2010ImageTransform,Sun2010TheWavelets,Prasad2018BookHwang}. Most notably, the Cohen–Daubechies–Feauveau-9/7 (CDF-9/7) wavelets \cite{Cohen1992BiorthogonalWavelets} are used as a key component of the JPEG2000 image compression standard \cite{Usevitch2001A2000,Skodras2001TheStandard,Taubman2002,Brislawn2003ResolutionStandard}. Similar to the well-established Fourier transform, DWTs can be applied to decompose a signal into different frequency components. However, as a form of multi-resolution analysis (MRA), wavelet transforms also resolve different length/time scales separately, thus are particular suited to analysis of features at different scales of observation and applications such as noise-reduction\cite{Mohideen2008ImageTransform,Chang2000AdaptiveCompression,Bnou2020AModel}.

\begin{figure} [!t]
    \centering
    \includegraphics[width=7.0cm]{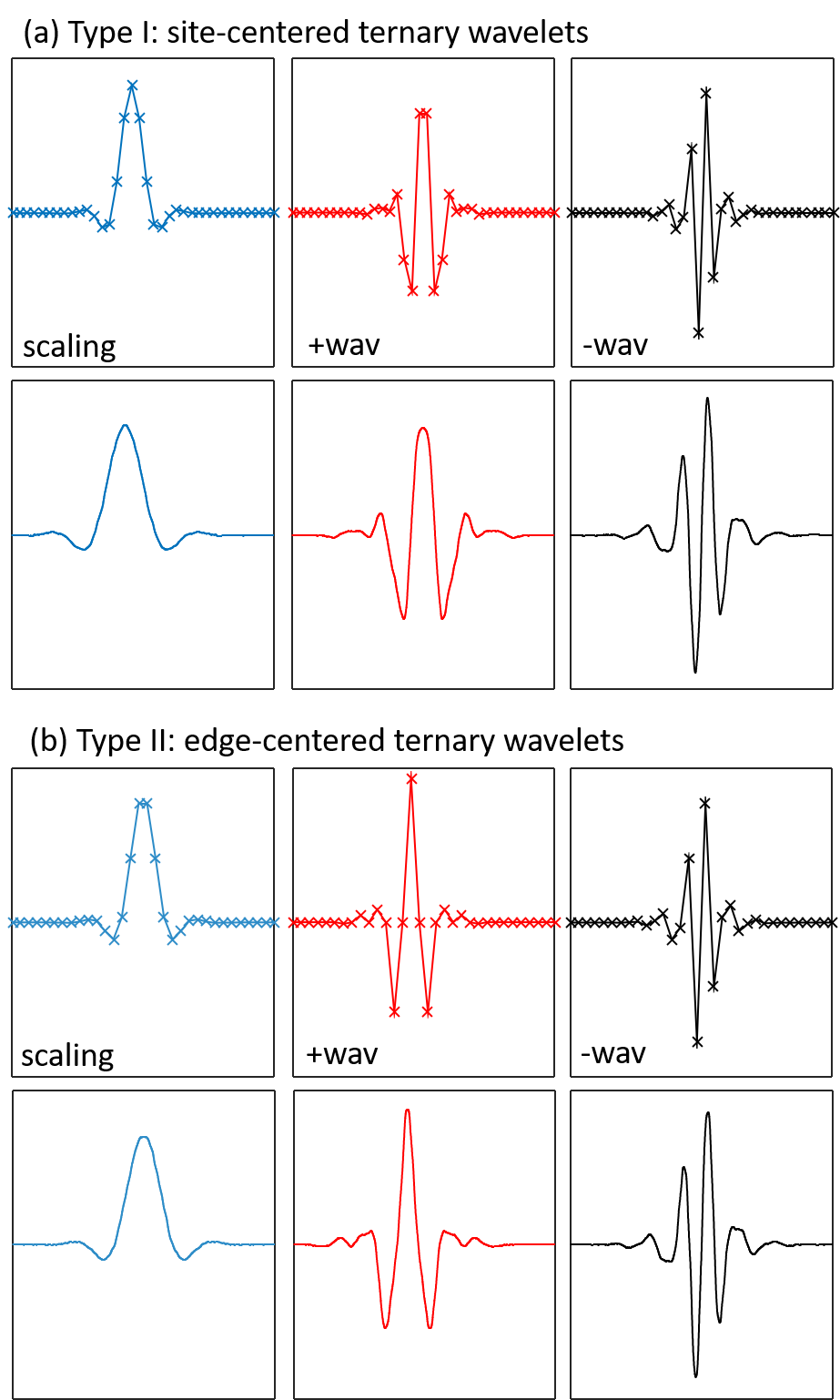}
    \caption{(a) Depictions of the scaling/wavelet sequences and the scaling/wavelet functions (in the continuum limit) associated to the Type-I ternary wavelets. From left-to-right, these consist of a (low-pass) symmetric scaling function, a (mid-pass) symmetric wavelet and a (high-pass) anti-symmetric wavelet. (b) Depictions of the scaling/wavelet sequences and the scaling/wavelet functions (in the continuum limit) associated to the Type-II ternary wavelets. Notice that the Type-I and Type-II functions have the similar symmetries, but that the scaling function from the former is symmetric about a single site while that from the latter is symmetric about an edge between two sites.}
    \label{fig:wavelets}
\end{figure}

It has long been known that the DWTs are related to forms of MRA used in many-body physics, such as the renormalization group \cite{Battle1999WaveletsRenormalization}. However, more recent works have established a precise connection between DWTs and a family of multi-scale unitary circuit \cite{Evenbly2016,Evenbly2018,Haegeman2018}, known as the multi-scale entanglement renormalization ansatz (MERA)\cite{Vidal2008}, used to describe quantum wavefunctions. Subsequently wavelets have also found wide utility towards quantum field theories and simulation algorithms for quantum many-body systems\cite{Singh2016HolographicWavelets, Michlin2017MultiresolutionBases, Lee2017GeneralizedWavelets, Fries2019RenormalizationWavelets, Polyzou2020WaveletTheory, Stottmeister2021Operator-AlgebraicWavelets, Morinelli2021ScalingWavelets,Witteveen2021BosonicTheory, Qiu2021HybridSets, George2022EntanglementRepresentations}. The establishment of a connection between wavelets and MERA networks opens an intriguing possibility: that the methodologies used to construct MERA may be exported to construct new families of DWTs. Such constructions were explored at length in Ref.\onlinecite{Evenbly2018}, where several families novel wavelet families were proposed. In this paper, we explore two of the most promising novel wavelets proposed in Ref.\onlinecite{Evenbly2018}, referred to as Type-I and Type-II ternary wavelets, for application towards image compression. Both of these wavelet types are based on a scale-3 (or dilation-3) transform, where each level represents a three-channel filter bank with well-defined low/mid/high pass filters, which are realized by a scaling function and a pair of wavelets. The wavelets we explore have many desirable properties: they are orthogonal, exactly reflection symmetric/anti-symmetric and they have high smoothness. It is for these reasons that we seek to test their effectiveness in a practical application, namely image compression. We compare against the CDF-9/7 wavelets, which are a well-established choice of wavelet for image and video compression.   
 
The manuscript is organised as follows. First we recap from Ref.\onlinecite{Evenbly2018} the essential the unitary circuit based formulation of the Type-I and Type-II ternary wavelets. Then we describe how symmetric boundary extensions can be implemented for the ternary wavelets, crucial to effectively treat data on open intervals such as images. Next, we describe the benchmark protocols used for testing compression efficiency and present the results of the benchmark tests performed on a variety of photo-sets, comprising several thousand images in total, when comparing the Type-I/Type-II ternary wavelets to the CDF-9/7 wavelets. Finally, we discuss the implications of these results and future directions for this research.

\section{Wavelets and Unitary Circuits}
\label{sec:waveletsCircuits}
Here, we briefly recap the formulation of DWTs as unitary circuits first described in Ref.\onlinecite{Evenbly2018}, which bears some similarity to the decomposition of wavelets into lifting steps\cite{Daubechies1998FactoringSteps}. A ubiquitous concept in field of quantum computation are \emph{unitary circuits}, where a unitary transformation acting on an extended lattice system are realized as a composition of multiple local unitary gates. In Ref.\onlinecite{Evenbly2018} it was argued that certain multi-scale unitary circuits, specifically those possessing a finite causal width\cite{Vidal2008}, can be understood as describing a system of orthogonal, compact wavelets. Note that, in this manuscript we consider circuits containing only real-valued (i.e. orthogonal) gates, but will continue to refer to them as unitary circuits none-the-less.

\begin{figure}
    \centering
    \includegraphics[width=8.5cm]{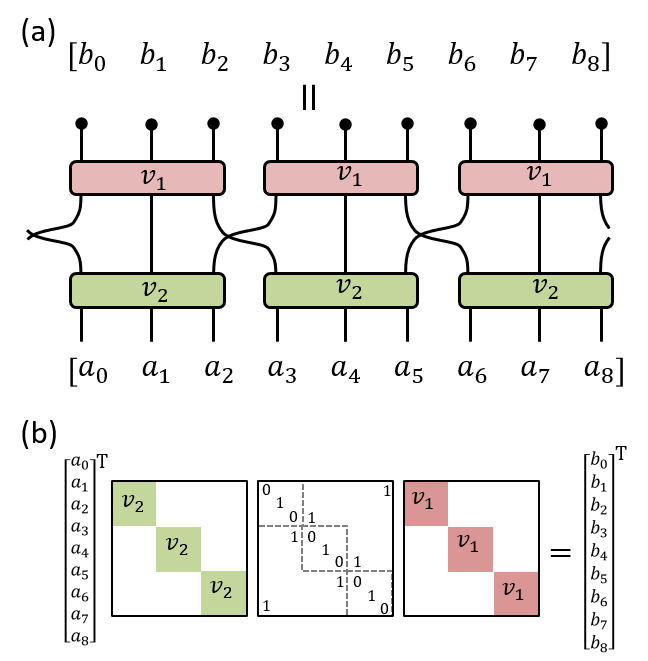}
    \caption{(a) A unitary circuit, which wraps periodically at the boundaries, maps an initial vector $\vec a$ of $N=9$ elements to a new vector $\vec b$. The circuit is here defined from a pair of $3$-body gates $\{v_1, v_2\}$ and copies thereof. Gates within the same row are interpreted as a direct sum of matrices, while different rows of the circuit are composed via matrix multiplication. The crossing of neighboring wires can equivalently be understood as enacting the permutation matrix $u_P$ of Eq.\ref{eq:up}. (b) An equivalent representation of the mapping $\vec a \rightarrow \vec b$ as a matrix equation.}
    \label{fig:circuit}
\end{figure}

A guide to interpreting circuit diagrams in the context of linear transforms and wavelets is presented in Fig.\ref{fig:circuit}. Here a gate with $n$ inputs/outputs is understood to represent an $n \times n$ unitary matrix, while a row of the circuit containing multiple gates (acting on different sites) represents a direct sum of the matrices, and vertically stacked gates or rows of the circuit are composed via matrix multiplication. Thus the unitary circuit acting on a system of $N$ sites describes a unitary transformation $U$ of this system or, equivalently, a basis of $N$ orthogonal functions for the system (i.e. corresponding to the columns of $U$). Notice that if a finite-depth the circuit is constructed with translation symmetry under shifts by $n$-sites, which can be achieved by constructing each row of the circuit as translations of an identical $n$-site gate, then the circuit can be understood to realize a single level of compactly-supported and orthogonal $n$-band wavelets.

\subsection{Ternary Unitary Circuits}
In this section we provide an overview of the ternary unitary circuits proposed in Ref.\onlinecite{Evenbly2018} which are used to construct the Type-I and Type-II ternary wavelets of Fig.\ref{fig:wavelets}. A core motivation for the proposal of ternary unitary circuits in Ref.\onlinecite{Evenbly2018} was to realize wavelets that were simultaneously orthogonal and reflection symmetric; we now recap the general strategy that was used to achieve this. A key aspect of the circuit formalism is that desired \emph{global} symmetries can be ensured by imposing \emph{local} symmetries on the individual gates comprising the circuit. Specifically, if one ensures that each gate is individually orthogonal and reflection symmetric, then the wavelets resulting from a composition of such gates will likewise be orthogonal and reflection symmetric. If we first consider $2\times 2$ matrices, then the only non-trivial orthogonal matrix with reflection symmetry is the permutation matrix $u_P$,
\begin{equation}
{u_P} = \left[ {\begin{array}{*{20}{c}}
0&1\\
1&0
\end{array}} \right]. \label{eq:up}
\end{equation}
Note that, in the context of circuit diagrams, we represent the permutation $u_P$ as a crossing of circuit wires rather than a distinct gate, see Fig.\ref{fig:circuit}. Clearly it is not possible to construct a viable wavelet transform from $u_P$ gates alone; it follows that larger gates must also be employed. If we now consider $3\times 3$ matrices then a family of orthogonal and reflection symmetric matrices is found,
\begin{equation}
 v_k = \frac{1}{2}\left[ {\begin{array}{*{20}{c}}
  {\cos \left( \theta_k  \right) + 1}&{\sqrt 2 \sin \left( \theta_k  \right)}&{\cos \left( \theta_k  \right) - 1} \\ 
  { - \sqrt 2 \sin \left( \theta_k  \right)}&{2\cos \left( \theta_k \right)}&{ - \sqrt 2 \sin \left( \theta_k  \right)} \\ 
  {\cos \left( \theta_k  \right) - 1}&{\sqrt 2 \sin \left( \theta_k  \right)}&{\cos \left( \theta_k  \right) + 1} 
\end{array}} \right] \hfill \label{eq:v}
\end{equation}
parameterized by a single angle $\theta_k \in [-\pi,\pi]$. 

\begin{figure}
    \centering
    \includegraphics[width=8cm]{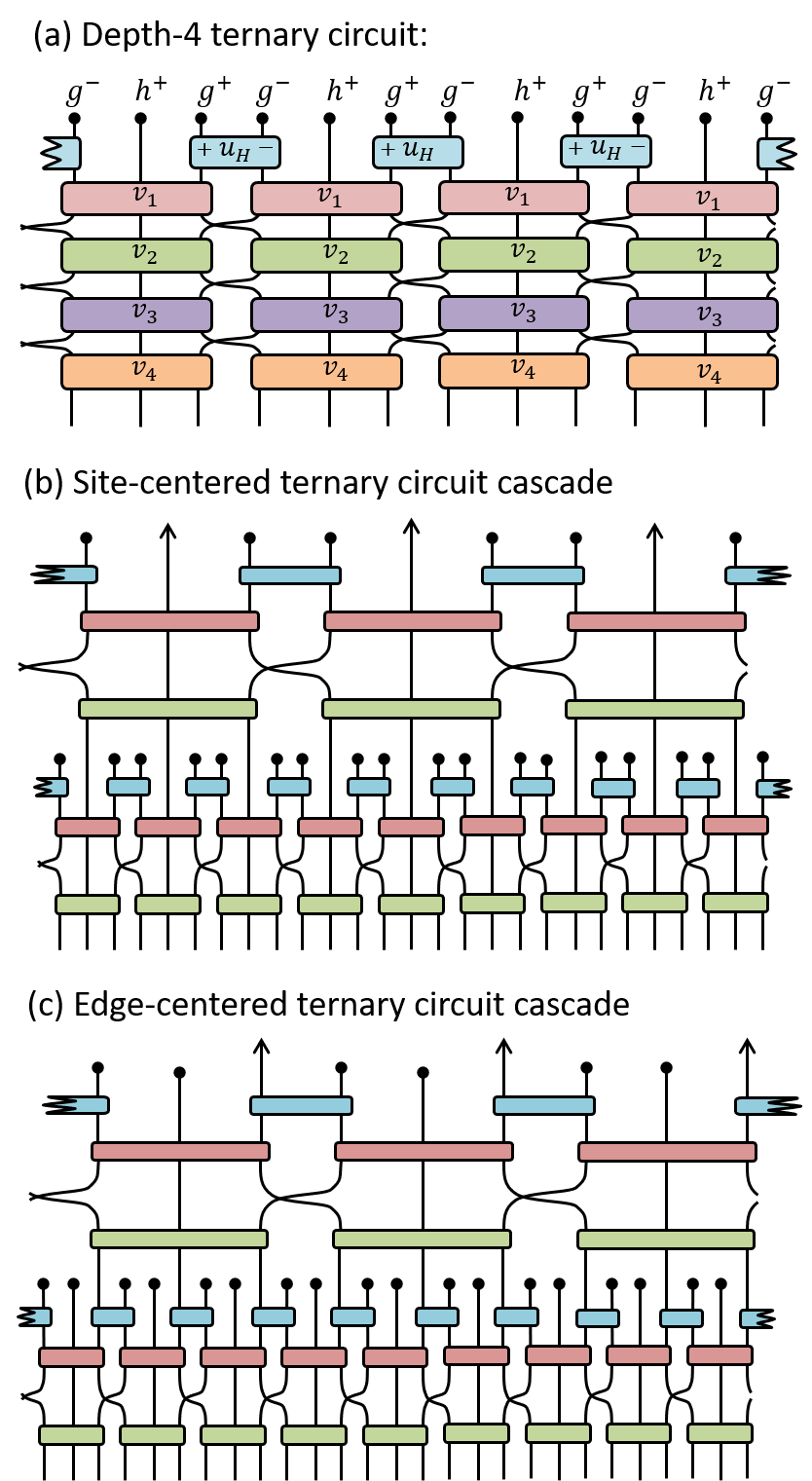}
    \caption{(a) A depth-4 ternary circuit, which effectively performs convolution of an input with a set of sequences $\{h^+, g^+, g^-$\}, is understood to represents a single level of a 3-band wavelet transform. The sequence $h^+$ is of length $N=21$ sites and is symmetric about the central site, while the pair of sequences $\{g^+, g^-\}$ are of length $N=24$ sites and are symmetric/anti-symmetric about the edge between their two central sites. (b) A site-centered composition of two depth-2 ternary circuits, which results from treating $h^+$ as the scaling sequence, represents two levels of the wavelet transform. (c) An edge-centered composition of two depth-2 ternary circuits, which results from treating $g^+$ as the scaling sequence, represents two levels of an alternative wavelet transform.}
    \label{fig:ternary}
\end{figure}

Ternary circuits are constructed from a particular composition of the $v_k$ matrices in conjunction with the $u_P$ permutation matrices. Specifically, a depth-$z$ ternary circuit is formed from $z$ distinct rows of 3-body $v_k$ gates, which are identical within each row but potentially different between separate rows, interspersed with permutation gates $u_P$ between these rows as depicted in Fig.\ref{fig:ternary}(a). At the top of each layer is an additional set of $2\times 2$ Hadamard gates,
\begin{equation}
{u_H} = \frac{1}{{\sqrt 2 }}\left[ {\begin{array}{*{20}{c}}
1&1\\
1&{ - 1}
\end{array}} \right] \label{eq:uh}
\end{equation}
which, although not reflection symmetric themselves, act to output a symmetric and anti-symmetric combination of their inputs, similar to their use in the Haar wavelets. Fig.\ref{fig:ternary}(a) depicts a depth-$4$ ternary circuit (i.e. containing 4 rows of $v_k$ gates), which is parameterized by a set of 4 angles $\{\theta_1, \theta_2, \theta_3, \theta_4\}$ which describe the gate angles of Eq.\ref{eq:v}. In general, ternary circuit of any arbitrarily depth $z\ge 1$ could be considered, which would similarly be parameterized by a set of $z$ gate angles.

Notice that any ternary circuit, regardless of its depth or the choice of angles parameterizing it, could be regarded as implementing one level of an orthogonal, symmetric wavelet transform. In particular the circuit implements convolution with a (site-centered) symmetric sequence $h^+$, with the associated coefficients output on the central index of $v_1$ gates, and with (edge-centered) symmetric/anti-symmetric sequences $\{g^+, g^-\}$, with the associated coefficients output from the $u_H$ gates. It is seen that for a depth-$z$ circuit the sequence $h^+$ has a length of $(6z-3)$ sites, while the sequences $\{g^+, g^-\}$ each have a length of $6z$ sites. Thus, while higher depth circuits contain more tuneable parameters $\theta_k$, this additional freedom comes at the expense of increasing the length of the associated scaling/wavelet sequences. 

Given that ternary circuits omit two distinct symmetric sequences, $h^+$ and $g^+$, there are also two distinct ways that one could cascade the ternary circuits in order to implement a multi-scale transform: (i) as shown in Fig.\ref{fig:ternary}(b) one could treat $h^+$ as the scaling sequence and $\{g^+, g^-\}$ as the wavelet sequences, which we refer to as the site-centered cascade, or (ii) as shown in Fig.\ref{fig:ternary}(c) one could treat $g^+$ as the scaling sequence and $\{h^+, g^-\}$ as the wavelet sequences, which we refer to as the edge-centered cascade. The Type-I and Type-II ternary wavelets considered in this manuscript, and depicted in Fig.\ref{fig:wavelets}, precisely arise from site-centered and edge-centered cascades, respectively, of depth-$6$ ternary circuits.

\subsection{Type-I/II Ternary Wavelets}
A key strength of the unitary circuit formalism for wavelets is that considerations of orthogonality/symmetry, as discussed in the previous section, have been entirely divorced from the consideration of resolving frequency bands, which we now discuss. In particular, given that the generic form of the ternary circuits from Fig.\ref{fig:ternary}(a) always preserve the desired symmetries (orthogonality and reflection symmetry), one is free to set the angles $\theta_k$ which parameterize the circuit without restriction. In this section we briefly describe the criteria from Ref.\onlinecite{Evenbly2018} for how the specific angles $\theta_k$ for the Type-I and Type-II ternary wavelets were chosen. 

The primary goal as stated in Ref.\onlinecite{Evenbly2018} was to set the angles such that the ternary circuits realized 3-band filters with well-resolved low/mid/high frequency components. Let us consider the Type-I (site-centered cascade) wavelets first, where the goal can be partly realized by ensuring that the two wavelet sequences $\{g_r^{+}, g_r^{+}\}$ have vanishing moments,
\begin{equation}
\sum\limits_{r} {\left( {r^\alpha {g_r^{+}}} \right)}  = 0, \phantom{xxx} \sum\limits_{r} {\left( {r^\alpha {g_r^{-}}} \right)}  = 0, \label{eq:vm1}
\end{equation}
for some sequence of values $\alpha$. While these constraints would be sufficient to ensure that the wavelets do not possess low-frequency components, additional constraints are also imposed to ensure that the scaling function $h_r^{+}$ and the symmetric wavelet $g_r^{+}$ do not possess high-frequency components. Specifically the sequences $h_r^{+}$ and $g_r^{+}$ are required to possess high-frequency `vanishing moments',
\begin{equation}
\sum\limits_{r} {\Big( {(-1)^r (r^\alpha) {g_r^{+}}} \Big)}  = 0, \phantom{xxx} \sum\limits_{r} {\Big( {(-1)^r (r^\alpha) {h_r^{+}}} \Big)}  = 0. \label{eq:vm2}
\end{equation}
for some sequence of values $\alpha$. The combined constraints of Eq.\ref{eq:vm1} and Eq.\ref{eq:vm2} can ensure that the sequences $\{h_r^{+}, g_r^{+}, g_r^{+}\}$ associated to the Type-I wavelets properly realize low/mid/high-pass frequency filers respectively. Similar constraints are also used for the Type-II (edge-centered cascade) wavelets, but with the roles of $g_r^{+}$ and $h_r^{+}$ sequences interchanged. 

For the case of depth-6 ternary circuits, a numeric optimization was used to find angles $\{\theta_1, \theta_2, \theta_3, \theta_4, \theta_5, \theta_6\}$ as given in Tab.\ref{tab:ternangles} which exactly satisfied (to within double-precision numerics) the constraints of Eq.\ref{eq:vm1} and Eq.\ref{eq:vm2} for values $\alpha=[0,1,2]$. These specific gate angles $\theta_k$, in conjunction with a site-centered or edge-centered cascade of the ternary circuit, define the Type-I and Type-II ternary wavelets respectively. It follows that, for both Type-I and Type-II constructions, the symmetric wavelets have 4 vanishing moments (as they trivially satisfy Eq.\ref{eq:vm1} for all odd $\alpha$) while the anti-symmetric wavelets have only 3 vanishing moments. Note that one can potentially optimise higher-depth circuits, which contain more free parameters, in order to satisfy the constraints of Eqs.\ref{eq:vm1} and \ref{eq:vm2} for a larger range of values $\alpha$. This would produce wavelets with more vanishing moments, though we do not consider this in the present work. Finally it is worth remarking that the Type II wavelets, although produced from depth-$6$ ternary circuits, contain only 3 non-trivial rotation angles, see Tab.\ref{tab:ternangles}, since gates $v_k$ with rotation angle $\theta_k = \pi$ simply reduce to (3-body) permutation matrices. This observation can be exploited to significantly reduce the computational cost of implementing a Type-II wavelet transform as further discussed in Sect.\ref{sect:comp}.

\begin{table}
\centering
\begin{tabular}{|c||c|c|}
\hline 
       & \phantom{xxxx} Type I \phantom{xxxx}  & \phantom{xxxx} Type II \phantom{xxxx}  \\ \hline
$\theta_1$ &  0.072130476  & -0.261582176   \\
$\theta_2$ &  0.847695078  & $\pi$          \\
$\theta_3$ & -0.576099009  & 0.107465734    \\
$\theta_4$ & -0.591746629  & $\pi$          \\
$\theta_5$ &  0.673886987  & -0.461363266   \\
$\theta_6$ &  0.529449713  & 0              \\
 \hline
\end{tabular}
\caption{Angles $\theta_k$ parameterizing the gates $v_k$ of Eq.\ref{eq:v} utilised for the Type-I and Type-II ternary wavelets.}
\label{tab:ternangles}
\end{table}

\section{Boundary Extensions}\label{sect:bound}
Most real-world photographic images, with the potential exception of something like a 360\textdegree\ panoramic image, should be understood as a finite signal with open boundaries, so for applications towards image compression it is important to understand how to efficiently apply a DWT to this case. A common approach to open boundaries is to `pad' the image at its boundaries (by an amount equal to the length of the scaling/wavelet sequences in use) and then to transform the padded image as usual. In the context of image compression several considerations should be made with respect to this padding: (i) the padding should be chosen to avoid discontinuities with the existing data, which will otherwise yield large many large wavelet coefficients, and (ii) the padding should avoid \emph{coefficient expansion}, where total number of scaling plus wavelet coefficients in the transformed signal is larger than the length of the original signal due. A significant advantage of exactly symmetric wavelets, such as the CDF wavelets, is that they can avoid the two aforementioned issues via a \emph{symmetric boundary extension}\cite{Brislawn1996ClassificationBanks, Usevitch2001A2000, Brislawn2007EquivalenceStandard}, which involves padding an image by mirroring the image across the boundary. Symmetric extensions are, by construction, always continuous across boundaries, while transformed image also retains the mirror symmetry (due to the symmetry of the DWT) such that the number of unique coefficients of the transformed image remains the same as the original image. In this section we describe how symmetric extensions are properly utilized for Type-I and Type-II ternary wavelets, the details of which are otherwise not obvious.

\begin{figure}
    \centering
    \includegraphics[width=8.5cm]{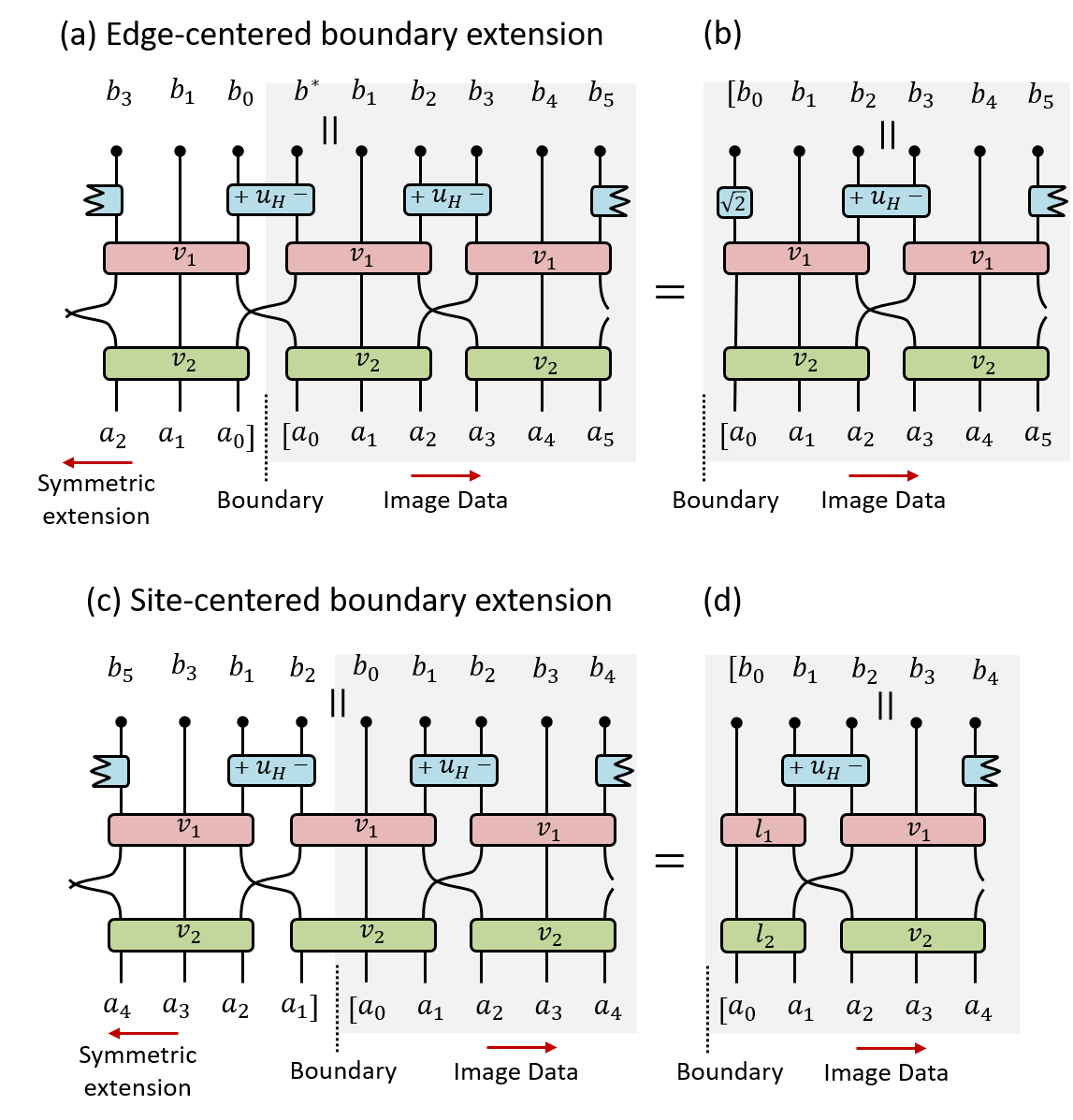}
    \caption{(a) A vector of input data $\vec a$ is mirrored across an edge at the left boundary and then mapped through the unitary circuit to give $\vec b$. The number of unique elements in $\vec b$ can be seen to match those in $\vec a$ given that the coefficient $b^*$, which corresponds to the product of an anti-symmetric wavelet with (symmetric) data across the boundary, is always exactly zero. (b) An open-boundary circuit that produces the same (unique) transformed data $\vec b$ as the symmetric extension from (a). (c) A vector of input data $\vec a$ is mirrored around a boundary site and then mapped through the unitary circuit to give $\vec b$. (d) An open-boundary circuit, with gates $l_k$ as defined in Eq.\ref{eq:lk}, reproduces the effect of the symmetric extension from (c).}
    \label{fig:boundary}
\end{figure}

We begin by noting that there are two different ways to symmetrically extend a finite signal that are compatible with the reflection symmetry of a ternary circuit. The first is to extend around an edge between two $v_k$ gates as shown Fig.\ref{fig:boundary}(a), which we refer to as an edge-centered extension. After performing this extension, one can see that most of the elements of the transformed data still mirror each other across the boundary, such that the duplicate coefficients do not need to be stored explicitly. However it initially appears that coefficient expansion is still necessary, as the transformed data contains an additional unique coefficient which we label $b^*$ in Fig.\ref{fig:boundary}(a). Yet, on closer inspection, it is seen that $b^* = 0$ is always exactly zero since this corresponds to product of anti-symmetric wavelet across a manifestly symmetric boundary, such that this coefficient can be safely ignored. At the practical level of implementing the DWT one could perform the symmetric extension explicitly (i.e. by generating a larger image with the appropriate padding in place), perform the transform, and then trim the transformed image at the boundary in order to retain only the unique coefficients. However, this explicit padding/trimming is not necessary. As seen in Fig.\ref{fig:boundary}(b), one can construct an open boundary circuit which perfectly mimics the effect of the symmetric extension without having to explicitly pad the data. In the present case this open boundary circuit is formed by simply removing the permutation gates $u_P$ at the boundary and adding a final $\sqrt 2$ weighting on the edge coefficient.

The second way of performing a symmetric extension for a ternary circuit is to extend from the center of $v_k$ gate as shown Fig.\ref{fig:boundary}(c), which we refer to as an site-centered extension. Again this avoids coefficient expansion, as the transformed signal has the same number of unique coefficients as that of the original signal. Here we can also avoid the need to explicitly pad the data by instead employing a properly chosen open boundary circuit that exactly mimics the effect of the symmetric extension, as seen in Fig.\ref{fig:boundary}(d). In this case we are required to introduce new 2-body gates $l_k$ on the left boundary,
\begin{equation}
{l_k} = \left[ {\begin{array}{*{20}{c}}
{\cos \left( {{\theta _k}} \right)}&{ - \sin \left( {{\theta _k}} \right)/\sqrt 2 }\\
{\sqrt 2 \sin \left( {{\theta _k}} \right)}&{\cos \left( {{\theta _k}} \right)}
\end{array}} \right] \label{eq:lk}
\end{equation}
which simply replicates the relevant outputs of the previously considered $v_k$ gates under a symmetric input. On the right boundary the corresponding 2-body gate would be ${r_k} = l_k^{\dag}$.

It is important to realize that the open boundary circuits of Fig.\ref{fig:boundary}(b) and  Fig.\ref{fig:boundary}(d) now technically represent a bi-orthogonal wavelet transformation since they contain non-orthogonal gates (although, as previously remarked, are still exactly equivalent to the orthogonal transformation applied to symmetrically extended signals). Given this consideration, it is apparent that the circuits corresponding to the inverse wavelet transforms are no longer just the transpose of forward transform circuit. For the case of the edge-centered open circuit in Fig.\ref{fig:boundary}(b) we introduce a $1/\sqrt 2 $ edge-weighting for inverse transform in place of the $\sqrt 2$ weighting. For the case of the site-centered open circuit in Fig.\ref{fig:boundary}(d) the left boundary gates $l_k$ should be replaced in the inverse circuit by new gates $\tilde l_k = (l_k^{\dag})^{-1}$, or explicitly
\begin{equation}
{\tilde l_k} = \left[ {\begin{array}{*{20}{c}}
{\cos \left( {{\theta _k}} \right)}&{ - \sqrt 2 \sin \left( {{\theta _k}} \right)}\\
{\sin \left( {{\theta _k}} \right)/\sqrt 2 }&{\cos \left( {{\theta _k}} \right)}
\end{array}} \right] \label{eq:ilk},
\end{equation}
while the right boundary gates are given as ${\tilde r_k} = \tilde l_k^{\dag}$

The motivation for introducing two distinct types of boundary extension, i.e. edge-centered and site-centered extensions, is that they can allow us to transform a signal of any length $N$ (i.e. not just where $N$ is a multiple of 3) by using an appropriate combination of left/right boundary extensions. As shown in Tab.\ref{tab:bound} one should choose the appropriate pair of extensions at left/right boundary based on the data length $N$ modulo 3. For instance, if $\bmod(N,3) = 0$ then edge-centered extensions should be performed at both left/right boundaries, as this is the only choice where the open-boundary circuit is of compatible length with $N$. For the case of $\bmod(N,3) = 2$ then either edge/site or site/edge would be viable options for extension at the left/right boundaries, however by convention we choose to exclude usage of the latter combination. Given these considerations, the sequence of appropriate left/right boundary extensions used for each level of the multi-scale wavelet transform are uniquely determined by the initial signal length $N_0$, such they do not need to be stored in order to later perform the inverse transformation.  

\begin{table}[]
\begin{tabular}{|c|c|c|c|c|}
\hline
\phantom{x} \#Data points \phantom{x}    &  L/R extension  & \phantom{x} \#Sca \phantom{x} & \phantom{x} \#Wav$+$ \phantom{x} & \phantom{x} \#Wav$-$ \phantom{x} \\ \hline
$N = 3k$    & edge / edge     & $k$      & $k+1$     & $k-1$     \\
$N = 3k +1$ & site / site     & $k+1$    & $k$       & $k$       \\
$N = 3k +2$ & edge / site     & $k+1$    & $k+1$     & $k$       \\ \hline
\end{tabular}
\caption{Conventions for the type of symmetric extensions utilised at the left/right edge of a signal based on the number $N$ of data points (with $k$ a positive integer). The labels \{\#Sca, \#Wav$+$, \#Wav$-$\} denote the number of scaling, symmetric wavelet and anti-symmetric wavelet coefficients produced following a single level of the transformation from the Type-I ternary wavelets. The Type-II ternary wavelets follow similarly, but with the \#Wav$+$ and \#Sca entries exchanged. } \label{tab:bound}
\end{table}
 
\section{Compression Benchmarks}
\label{sect:methods}
\subsection{Protocol}
We now describe the benchmark protocol used for comparing the performance of the Type-I and Type-II ternary wavelets against the CDF-9/7 wavelets. Our protocol is based on comparing the minimum number of wavelet coefficients $M$ needed to achieve a preset image quality as quantified by the multi-scale structural similarity index measure (MS-SSIM) \cite{Wang2004ImageSimilarity.}, which is widely regarded as a good metric for determining the change in perceived image quality \cite{Wang2004ImageSimilarity.,Gore2015FullImages,ZhouWang2011InformationAssessment,Zhang2012AAlgorithms,Sgaard2016ApplicabilityStreaming}. We test each wavelet encoding over a range of qualities, specifically with fixed MS-SSIM = $[0.99, 0.98, 0.95]$, which roughly correspond to high/medium/low quality encodings respectively.

Benchmark comparisons are performed using the University of Washington Groundtruth Database \cite{Shapiro2004UniversityDatabase}, consisting of 845  color photo images each around 0.4-megapixels in resolution, the Uncompressed Colour Image Database (UCID) \cite{ucidDatabase}, consisting of 1338 uncompressed color photo images each around 0.2-megapixels in resolution. Additionally, in order to obtain results more representative of photographs produced by modern digital cameras, we also benchmark using a selection of 100 high-resolution photo images, each 16 to 40-megapixels in resolution, taken randomly from the landscapes/people/animals section of a photography website and available via Ref.\onlinecite{ternCode2022}.

The protocol for evaluating each type of wavelet on a single image is as follows: (i) the image is transformed to the wavelet basis (applying as many recursive levels of wavelet expansion as viable given the image size), (ii) the wavelet coefficients are thresholded to keep the only the $M$ largest wavelet coefficients while zero-ing out the remaining coefficients, (iii) the inverse wavelet transform is performed such that an approximation to the original image is recovered and (iv) the MS-SSIM is computed between the original and the approximate image. The steps (ii) though (iv) are performed multiple times, adjusting the number of coefficients $M$ kept, until the minimum $M$ meeting the prescribed quality requirement is determined. For each image we the compare the minimum required coefficients from the ternary wavelets $M_\textrm{tern}$ against the minimum required coefficients from the CDF-9/7 wavelets $M_\textrm{CDF}$, and then calculate the relative compression performance $\beta_c$,
\begin{equation}
\beta_c = 1 - \frac{{{M_\textrm{tern}}}}{{{M_\textrm{CDF}}}} \label{eq:rel}
\end{equation}

The CDF-9/7 wavelet transforms were performed using code derived from their implementation in the MATLAB Wavelet Toolbox\texttrademark, and the Type-I and Type-II ternary wavelets were implemented via custom MATLAB code available via Ref.\onlinecite{ternCode2022}. In all cases the images were pre-processed into YCbCr colorspace and employed symmetric boundary extensions. The MS-SSIM was computed using the `multissim` function from the MATLAB Image Processing Toolbox\texttrademark with default options. It should be noted that our testing protocol for image compression performance omits many important steps that are present in a fully-realized compression routine, such as the JPEG2000 standard\cite{Skodras2001TheStandard, Taubman2002}, including zero-tree embedding\cite{Shapiro1993EmbeddedCoefficients} and quantization\cite{Bradley1994Wavelet/scalarImages}. These addition steps would be required in order to test the actual compression performance in a realistic application. However, given that the additional steps would follow similarly regardless of the specific wavelets in use, there is no reason to believe that they would substantially affect the relative compression performance evaluated using our protocol. 


\begin{figure}
    \centering
    \includegraphics[width=8.5cm]{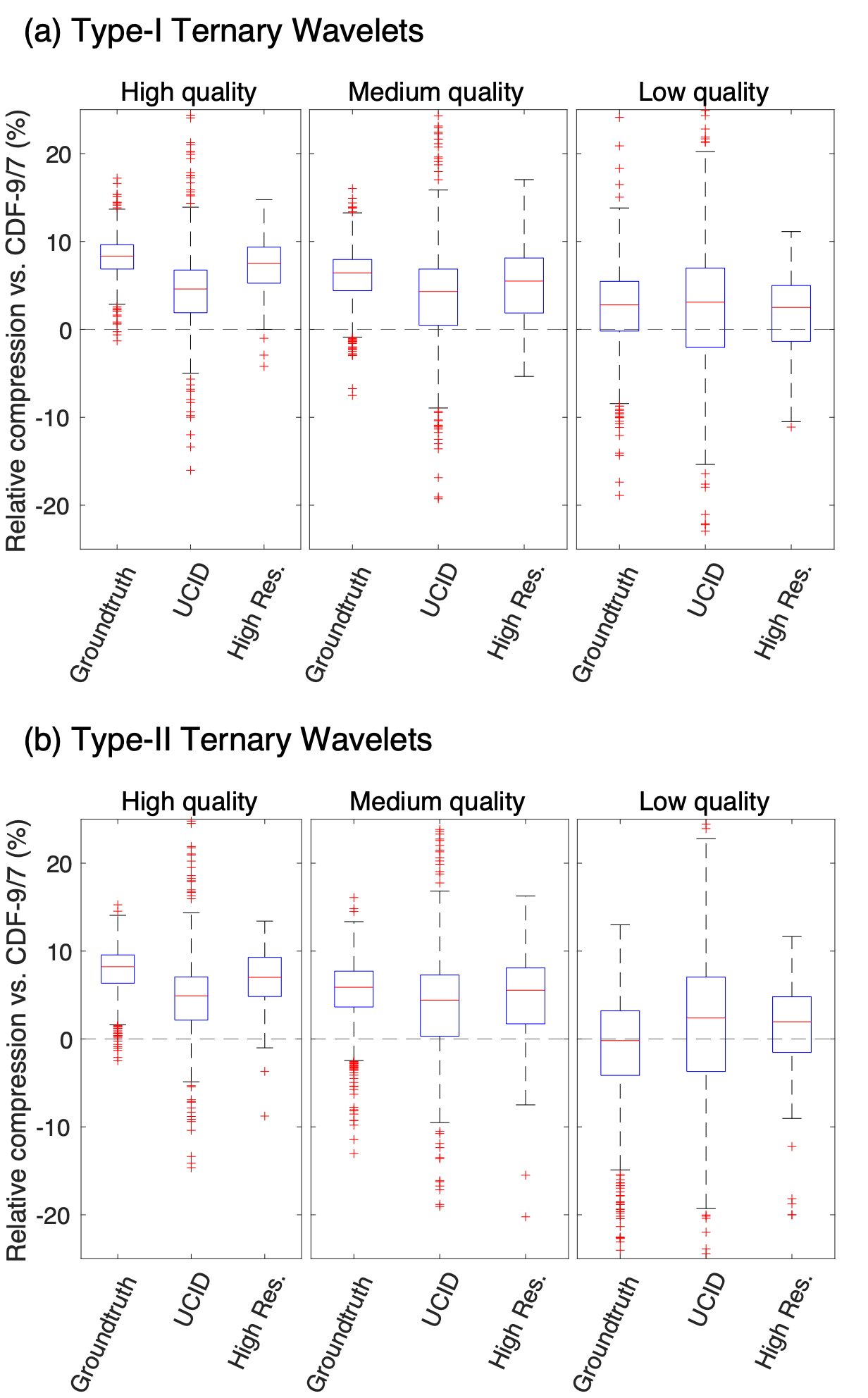}
    \caption{(a) Relative compression efficiency, as defined in Eq.\ref{eq:rel}, of the Type-I ternary wavelets versus the CDF-9/7 wavelets with high/medium/low quality encodings (corresponding to MS-SSIM fixed at $[0.99,0.98,0.95]$ respectively) of Groundtruth, UCID and high-res photo data-sets. The Type-I wavelets are seen to offer more efficient compression (on average) than the CDF-9/7 wavelets, with the performance gap widening at higher-quality encodings. For instance, the Type-I wavelets yielded a median reduction of $8.3\%$ fewer coefficients than the CDF-9/7 wavelets when encoding the Groundtruth data-set at high quality (MS-SSIM=0.99), while only yielding a median reduction of $2.8\%$ fewer coefficients at low quality (MS-SSIM=0.95). (b) Equivalent data for the Type-II ternary wavelets versus the CDF-9/7 wavelets. }
    \label{fig:results}
\end{figure}

\subsection{Results}
\label{sect:results}
The benchmarking results are shown in Fig.\ref{fig:results}, comparing the relative performance $\beta_c$ from Eq.\ref{eq:rel} of the ternary wavelets versus the CDF-9/7 wavelets at a fixed qualities of MS-SSIM = $[0.99,0.98,0.95]$. The box and whisker plots show the median, $25^{th}/75^{th}$ percentiles, and maximum/minimum values for tests run in that category of images. The outliers are plotted as red crosses and are defined as x1.5 the interquartile range from the $25^{th}/75^{th}$ percentiles. The results for the median relative performance are also summarised in Table.\ref{tab:compressionPerformanc}. While the relative compression performance varied between the different photo-sets and the different compression qualities, it is seen that both the Type-I and Type-II ternary wavelets consistently outperform the CDF-9/7 wavelets. Indeed, it is only for the specific case of the Groundtruth photo-set at low quality encoding (MS-SSIM=0.95) that the median performance of the Type-II wavelets is $0.2\%$ worse than the CDF-9/7 wavelets, with improved median performance showing in all other instances. 

It is interesting to note in Fig.\ref{fig:results} that the performance advantage of the ternary wavelets is greater when restricted to a higher quality setting (MS-SSIM=0.99) than when restricted to a lower quality setting (MS-SSIM=0.95). At the high quality setting this advantage is substantial: the Type-I and Type-II ternary wavelets require a median of $7-8\%$ fewer coefficients than the CDF-9/7 to compress in image from the Groundtruth or High-res photo-sets to achieve the same MS-SSIM. At the low quality setting the advantage is much reduced ($0-3\%$) although still potentially significant. The compression advantage of the ternary wavelets was also observed to be lesser for the UCID photo-set which, at around 0.2-megapixels per photo, contained lower resolution photos than the other photo-sets. Finally, it is worth remarking that only a slight difference between the performance of the Type-I and Type-II wavelets was observed, which is not surprising given the similarly between these ternary wavelets, with the Type-I wavelet offering a small performance gain across all categories.  


\begin{table}[]
\begin{tabular}{|c||c||c|c|c|}
\hline
                         & MS-SSIM & Groundtruth & \phantom{x} UCID \phantom{x} & \phantom{x} High-res \phantom{x} \\ \hline
\multirow{3}{*}{\shortstack[c]{Type-I Ternary\\vs CDF-9/7}}  & 0.99    & 8.3\% & 4.6\%       & 7.5\%    \\
                         & 0.98    & 6.4\% & 4.3\%       & 5.5\%    \\
                         & 0.95    & 2.8\% & 3.1\%       & 2.5\%    \\ \hline
\multirow{3}{*}{\shortstack[c]{Type-II Ternary\\vs CDF-9/7}} & 0.99    & 8.2\% & 4.9\%       & 7.0\%    \\
                         & 0.98    & 5.9\% & 4.4\%       & 5.5\%    \\
                         & 0.95    & -0.2\% &2.4\%      & 2.0\%    \\ \hline
\end{tabular}
\caption{\label{tab:compressionPerformanc}Median values of the relative compression performance $\beta_c$, see Eq.\ref{eq:rel}, of Type-I and Type-II ternary wavelets versus the CDF-9/7 wavelets at different quality requirements (specified by a fixed MS-SSIM) and across three image databases (Groundtruth, UCID, High-res). In all cases, except for the Groundtruth set at MS-SSIM=0.95 with the Type-II wavelets, the ternary wavelets have improved median compression relative to the CDF-9/7 wavelets.}
\end{table}

\section{Discussion}
\subsection{Compression Performance}
Previous explorations of compression efficiency with (different) ternary wavelets\cite{Zhou2006ConstructionFrame} have not yielded substantive advantages over the CDF-9/7 wavelets. It remains an interesting question why the Type-I and Type-II ternary wavelets have superior compression performance over CDF-9/7, especially given that a naive comparison of their properties might suggest otherwise. In particular the CDF-9/7 possess more vanishing moments (VMs) than the ternary wavelets; with 4 VMs for the former versus either 4VMs/3VMs for the latter (for the symmetric/anti-symmetric wavelets respectively). Additionally the CDF-9/7 wavelets are significantly more compact in terms of the lengths of their scaling/wavelet sequences, at less than one-third the length of those from the Type-I and Type-II ternary wavelets. We speculate that, while the ternary wavelets have fewer vanishing moments than the CDF-9/7 wavelets, their higher degree of smoothness could be the reason why the coefficient thresholding used in the benchmark protocol has less of an impact on the perceived image quality. It is also worth noting, although the scaling/wavelet sequences associated to the ternary wavelets are indeed long (at either 33 or 36 sites), that the coefficients only have significant weight on a small number of sites such that they \emph{effectively} remain quite compact. 

\begin{figure}
    \centering
    \includegraphics[width=8.5cm]{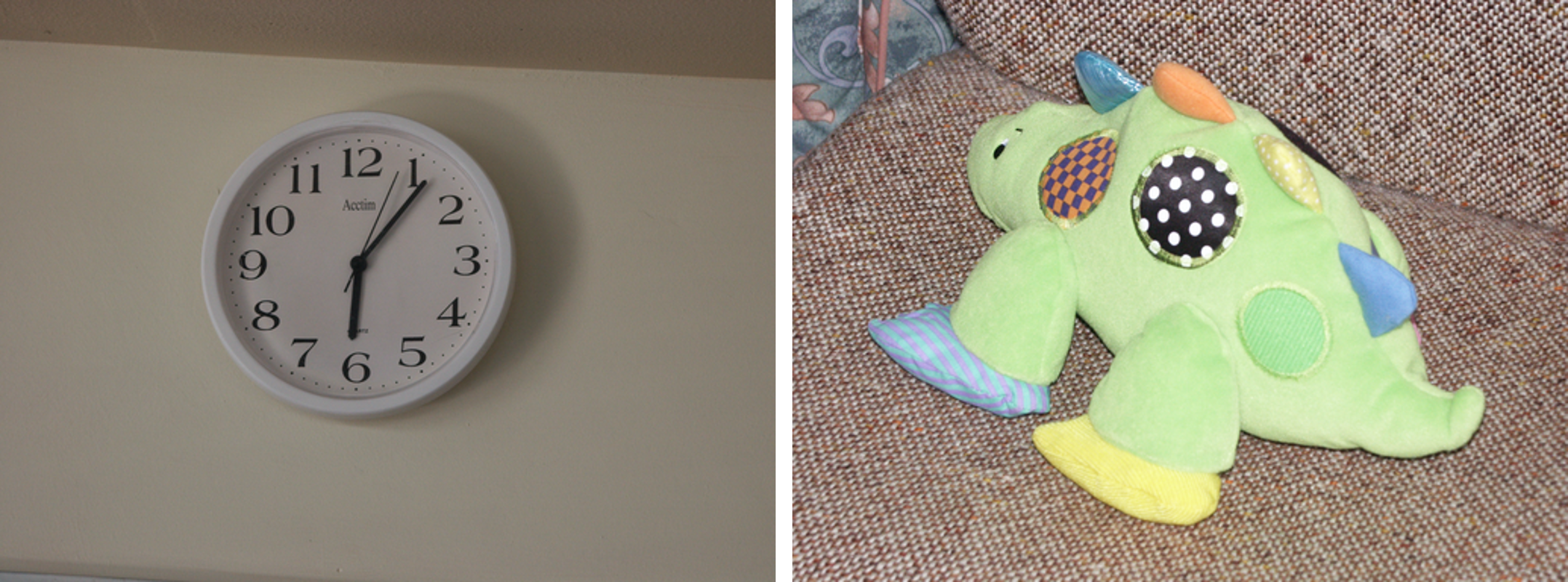}
    \caption{Example images from the UCID database where the ternary wavelets (left) under-performed and (right) over-performed the CDF-9/7 wavelets by substantial margins. Other instances where the ternary wavelets under-performed were similarly associated to low-detail photos.}
    \label{fig:UCID}
\end{figure}

Further hints regarding the origin of the performance differences can be gleaned by examining some specific instances of photos with either under-performing or over-performing compression from the Ternary wavelets. As per the examples in Fig.\ref{fig:UCID}, we find that most of the examples where the ternary wavelets over-performed (i.e. in the top $25^\textrm{th}$ performance percentile for relative compression) were typically images with high detail, such as those with textured backdrops. Conversely, the examples where the ternary wavelets under-performed (i.e. bottom $25^\textrm{th}$ percentile) were typically images with low detail, such as those with a small subject on an almost uniform background, which required a relatively small number of wavelet coefficients to meet the desired quality thresholds. 

\subsection{Computational Efficiency} \label{sect:comp}
While the primary goal of this manuscript was to explore the compression efficiency of the ternary wavelets as compared to the CDF-9/7 wavelets, it is also relevant to compare the computational costs of performing the DWTs. For purposes of this comparison we evaluate the total number of scalar multiplications $\mu$ required to perform each DWT, which serves as a rough proxy to the full computational cost.

We begin by recapping the cost of implementing a DWT based on the CDF-9/7 wavelets. Given a length-$N$ signal, the cost of applying a single level of the CDF-9/7 wavelet transform would require $\mu = 8N$ scalar multiplications when using a convolution based approach\cite{Cohen1992BiorthogonalWavelets}, but this could be reduced to $\mu = 3N$ by factoring into appropriate lifting steps\cite{Daubechies1998FactoringSteps}. It follows that applying a single level of the lifting-based transform on (the rows and columns of) an $N \times N$ image would require $\mu = 6N^2$, while transforming through all $\log_2 (N)$ levels would require a total of $\mu = 8N^2$ operations.  

Next we analyze the cost of performing the DWT associated to the Type-I and Type-II ternary wavelets. Given a length-$N$ signal, the cost of applying a single row of the 3-body gates $v_k$ would naively resolve to $\mu = 3N$ scalar multiplications. However, one could scale each gate $v_k$ by a factor $\sqrt 2 / \sin(\theta_k)$, such that four of the matrix elements in the re-scaled gate are of unit magnitude, which reduces the number of multiplications needed. The cost of applying a single row of (re-scaled) $v_k$ gates is then given as $\mu = (5/3)N$, such that the cost of applying the depth-6 ternary circuit evaluates as $\mu = 10N$. It follows that applying a single level of the ternary wavelet transform on an $N \times N$ image would require $\mu = 20N^2$, while transforming through all levels would require a total of $\mu = 22.5N^2$ operations. However, it should be noted, as evidenced by Tab.\ref{tab:ternangles}, that three of the gates $v_k$ required for the Type-II wavelet transform require no multiplications as they are either the identity matrix or a permutation matrix. Thus the cost of applying the Type-II wavelet transform is equivalent to that of a depth-3 ternary circuit, such that it is only half of the cost of the Type-I wavelet transform, or about $\mu = 11.25N^2$ multiplications for the transformation of an $N \times N$ image. 

In summary, we estimate that performing a wavelet transform with the Type-I ternary wavelets is almost 3 times more computationally expensive than with the CDF-9/7 wavelets. This would leave the Type-I wavelets at a serious disadvantage in applications where speed of compression/decompression is prioritized over compression efficiency. The Type-II ternary wavelets were found to be 1.4 times (or $40\%$) more computationally expensive than CDF-9/7 wavelets which, while still an appreciable increase in cost, is still greatly more manageable than the cost increase from the Type-I ternary wavelets. These estimates of the computational costs roughly coincide with the actual computation times observed in practice during the evaluation of the results. Finally, we remark that computational efficiency was certainly not the primary consideration in our proposal for the use of Type-I/Type-II ternary wavelets for image compression. Several of the other families of wavelets proposed in Ref.\onlinecite{Evenbly2018}, or from other unitary circuits specifically designed to minimize the computational cost, may be more competitive with the CDF-9/7 wavelets in this aspect.

\section{Conclusions}
\label{sect:conclusions}
In this manuscript we have described some of the implementation details of applying the Type-I and Type-II ternary wavelets proposed in Ref.\onlinecite{Evenbly2018} for applications in image compression, and tested their compression performance relative to the CDF-9/7 wavelets. The ternary wavelets, which realize a 3-band filter, were chosen for this comparison as they have an appealing set of features: they are orthogonal (i.e. energy conserving) while also possessing exact reflection symmetry, allowing open boundaries to be easily treated via symmetric extensions. The wavelets are also relatively smooth and are effectively quite compact. 
When benchmarked on a variety of image photo-sets, comprising several thousands of photos in total, both the Type-I and Type-II ternary wavelets are found to appreciably outperform the CDF-9/7 wavelets in terms of compression efficiency, with the improvements observed across a broad range of image resolutions and quality requirements. This improved efficiency of would allow images to be stored more compactly, while maintaining the same perceived quality as specified by the MS-SSIM. We speculate that the ternary wavelets might be especially useful if applied to video compression giving the modern prevalence of video streaming\cite{Bhaskaran1997ImageArchitectures, Prasad2018BookHwang}, as they could potentially reduce in the bandwidth required for video streaming as compared to CDF-9/7 based compression, or for other applications such as medical imaging\cite{Dhankhar2021AnalysisImages} or compression for large-scale climate data\cite{Woodring2011RevisitingPrecision}.
 
The results presented in this manuscript also highlight some of the advantages of the unitary circuit formalism for constructing wavelets presented in Ref.\onlinecite{Evenbly2018}. In particular, the circuit based constructions allow the properties relating to wavelet structure and symmetry (i.e. the length of scaling/wavelet sequences, orthogonality, reflection symmetry) to be considered separately from the frequency resolving properties (i.e. vanishing moments). This follows as the former properties are imposed through the choice of circuit and restrictions on the circuit gates, while the latter properties are set through tuning of the free parameters within the gates. Finally, we remark that the circuit formalism may be useful in the construction of more highly-optimized wavelets. Although the Type-I and Type-II ternary wavelets used in this manuscript were built according to some standard criteria (i.e. minimizing vanishing moments of wavelets), the circuit based formalism could be fully exploited to directly optimise wavelets for maximal efficiency. This could be accomplished, for instance, by tuning the angles $\theta_k$ of the circuit gates in Eq.\ref{eq:v} to minimise the loss at a prescribed compression ratio given a trial data set. Explorations in this direction remains an interesting avenue for future research.

The authors thank Steve R White for useful comments and discussion regarding this work.


\end{document}